\input harvmac
\input epsf.tex
\def\N{{\cal N}}

\noblackbox
%
% dual defs
%
%  \slashchar puts a slash through a character to represent contraction
 %  with Dirac matrices.
 \def\slash#1{\setbox0=\hbox{$#1$}           % set a box for #1
 \dimen0=\wd0                                 % and get its size
 \setbox1=\hbox{/} \dimen1=\wd1               % get size of /
 \ifdim\dimen0>\dimen1                        % #1 is bigger
 \rlap{\hbox to \dimen0{\hfil/\hfil}}      % so center / in box
  #1                                        % and print #1
 \else                                        % / is bigger
 \rlap{\hbox to \dimen1{\hfil$#1$\hfil}}   % so center #1
 /                                         % and print /
 \fi}                                         %

\def\ev#1{\langle#1\rangle}

%\IntriligatorJJ

\lref\thooftr{G. 't Hooft, in {\it Recent Developments in Gauge Theories}, eds. G. 't Hooft et. al.
(Plenum Press, New York, 1980) 135.}
\lref\IWbar{
K.~Intriligator and B.~Wecht,
``Baryon charges in 4D superconformal field theories and their AdS  duals,''
Commun.\ Math.\ Phys.\  {\bf 245}, 407 (2004)
[arXiv:hep-th/0305046].
%%CITATION = HEP-TH 0305046;%%
}
%\BarnesJJ
\lref\BarnesJJ{
  E.~Barnes, K.~Intriligator, B.~Wecht and J.~Wright,
  ``Evidence for the strongest version of the 4d a-theorem, via a-maximization
  along RG flows,''
  Nucl.\ Phys.\ B {\bf 702}, 131 (2004)
  [arXiv:hep-th/0408156].
  %%CITATION = HEP-TH 0408156;%%
}

%\BrodieVV
\lref\BCI{
J.~H.~Brodie, P.~L.~Cho and K.~A.~Intriligator,
``Misleading anomaly matchings?,''
Phys.\ Lett.\ B {\bf 429}, 319 (1998)
[arXiv:hep-th/9802092].
%%CITATION = HEP-TH 9802092;%%
}

%\IntriligatorRX
\lref\ISS{
K.~A.~Intriligator, N.~Seiberg and S.~H.~Shenker,
``Proposal for a simple model of dynamical SUSY breaking,''
Phys.\ Lett.\ B {\bf 342}, 152 (1995)
[arXiv:hep-ph/9410203].
%%CITATION = HEP-PH 9410203;%%
}

%\ErdmengerYC
\lref\JEHO{
  J.~Erdmenger and H.~Osborn,
  ``Conserved currents and the energy-momentum tensor in conformally  invariant
  theories for general dimensions,''
  Nucl.\ Phys.\ B {\bf 483}, 431 (1997)
  [arXiv:hep-th/9605009].
  %%CITATION = HEP-TH 9605009;%%
}

\lref\JJN{
I.~Jack, D.~R.~T.~Jones and C.~G.~North,
``$N=1$ supersymmetry and the three loop anomalous dimension for the chiral
superfield,''
Nucl.\ Phys.\ B {\bf 473}, 308 (1996)
[arXiv:hep-ph/9603386];
%%CITATION = HEP-PH 9603386;%%
I.~Jack, D.~R.~T.~Jones and C.~G.~North,
``Scheme dependence and the NSVZ beta-function,''
Nucl.\ Phys.\ B {\bf 486}, 479 (1997)
[arXiv:hep-ph/9609325].
%%CITATION = HEP-PH 9609325;%%
}
\lref\KutasovXU{
  D.~Kutasov and A.~Schwimmer,
  ``Lagrange multipliers and couplings in supersymmetric field theory,''
  Nucl.\ Phys.\ B {\bf 702}, 369 (2004)
  [arXiv:hep-th/0409029].
  %%CITATION = HEP-TH 0409029;%%
}
%\AnselmiRD
\lref\Anselmi{
  D.~Anselmi,
  ``Central functions and their physical implications,''
  JHEP {\bf 9805}, 005 (1998)
  [arXiv:hep-th/9702056].
  %%CITATION = HEP-TH 9702056;%%
}
%\IntriligatorEX
\lref\ISmir{
  K.~A.~Intriligator and N.~Seiberg,
  ``Mirror symmetry in three dimensional gauge theories,''
  Phys.\ Lett.\ B {\bf 387}, 513 (1996)
  [arXiv:hep-th/9607207].
  %%CITATION = HEP-TH 9607207;%%
}
%\AharonyBX
\lref\ISmirr{
  O.~Aharony, A.~Hanany, K.~A.~Intriligator, N.~Seiberg and M.~J.~Strassler,
  ``Aspects of N = 2 supersymmetric gauge theories in three dimensions,''
  Nucl.\ Phys.\ B {\bf 499}, 67 (1997)
  [arXiv:hep-th/9703110].
  %%CITATION = HEP-TH 9703110;%%
}
\lref\GW{D.~J.~Gross and F.~Wilczek,
``Asymptotically Free Gauge Theories. 2,''
Phys.\ Rev.\ D {\bf 9}, 980 (1974).
%%CITATION = PHRVA,D9,980;%%
}

\lref\BZ{T.~Banks and A.~Zaks,
``On The Phase Structure Of Vector - Like Gauge Theories With Massless Fermions,''
Nucl.\ Phys.\ B {\bf 196}, 189 (1982).
%%CITATION = NUPHA,B196,189;%%
}

%\AnselmiXK
\lref\AnselmiXK{
  D.~Anselmi,
  ``Anomalies, unitarity, and quantum irreversibility,''
  Annals Phys.\  {\bf 276}, 361 (1999)
  [arXiv:hep-th/9903059].
  %%CITATION = HEP-TH 9903059;%%
}
\lref\KINS{K.~A.~Intriligator and N.~Seiberg,
``Phases of N=1 supersymmetric gauge theories in four-dimensions,''
Nucl.\ Phys.\ B {\bf 431}, 551 (1994)
[arXiv:hep-th/9408155].
%%CITATION = HEP-TH 9408155;%%
}

%\SeibergBZ
\lref\SeibergBZ{
  N.~Seiberg,
  ``Exact results on the space of vacua of four-dimensional SUSY gauge
  theories,''
  Phys.\ Rev.\ D {\bf 49}, 6857 (1994)
  [arXiv:hep-th/9402044].
  %%CITATION = HEP-TH 9402044;%%
}
\lref\NSd{
N.~Seiberg,
``Electric - magnetic duality in supersymmetric nonAbelian
gauge theories,''Nucl.\ Phys.\ B {\bf 435}, 129
(1995)[arXiv:hep-th/9411149].
%%CITATION = HEP-TH 9411149;%%}
}
%\IntriligatorAU
\lref\ISrev{
K.~A.~Intriligator and N.~Seiberg,
``Lectures on supersymmetric gauge theories and electric-magnetic  duality,''
Nucl.\ Phys.\ Proc.\ Suppl.\  {\bf 45BC}, 1 (1996)
[arXiv:hep-th/9509066].
%%CITATION = HEP-TH 9509066;%%
}
%\OsbornQU
\lref\HOsc{
  H.~Osborn,
  ``N = 1 superconformal symmetry in four-dimensional quantum field theory,''
  Annals Phys.\  {\bf 272}, 243 (1999)
  [arXiv:hep-th/9808041].
  %%CITATION = HEP-TH 9808041;%%
}

%\AppelquistRB
\lref\AppelquistRB{
  T.~Appelquist, A.~Ratnaweera, J.~Terning and L.~C.~R.~Wijewardhana,
  %``The phase structure of an SU(N) gauge theory with N(f) flavors,''
  Phys.\ Rev.\ D {\bf 58}, 105017 (1998)
  [arXiv:hep-ph/9806472].
  %%CITATION = HEP-PH 9806472;%%
}
%\BrodieVV
\lref\BCI{
  J.~H.~Brodie, P.~L.~Cho and K.~A.~Intriligator,
  ``Misleading anomaly matchings?,''
  Phys.\ Lett.\ B {\bf 429}, 319 (1998)
  [arXiv:hep-th/9802092].
  %%CITATION = HEP-TH 9802092;%%
}
%\LatorreEA
\lref\LatorreEA{
  J.~I.~Latorre and H.~Osborn,
  ``Modified weak energy condition for the energy momentum tensor in  quantum
  field theory,''
  Nucl.\ Phys.\ B {\bf 511}, 737 (1998)
  [arXiv:hep-th/9703196].
  %%CITATION = HEP-TH 9703196;%%
}
 
%\OsbornCR
\lref\HOAP{
  H.~Osborn and A.~C.~Petkou,
  ``Implications of conformal invariance in field theories for general
  dimensions,''
  Annals Phys.\  {\bf 231}, 311 (1994)
  [arXiv:hep-th/9307010].
  %%CITATION = HEP-TH 9307010;%%
}

%\DuffWM
\lref\DuffWM{
  M.~J.~Duff,
  ``Twenty years of the Weyl anomaly,''
  Class.\ Quant.\ Grav.\  {\bf 11}, 1387 (1994)
  [arXiv:hep-th/9308075].
  %%CITATION = HEP-TH 9308075;%%
}

%  draw box of size #1pt and line thickness #2pt
\def\drawbox#1#2{\hrule height#2pt
             \hbox{\vrule width#2pt height#1pt \kern#1pt \vrule
width#2pt}
                   \hrule height#2pt}
% Young tableaux

\def\Fund#1#2{\vcenter{\vbox{\drawbox{#1}{#2}}}}
\def\Asym#1#2{\vcenter{\vbox{\drawbox{#1}{#2}
                   \kern-#2pt       % line up boxes
                   \drawbox{#1}{#2}}}}

\def\asym{\Asym{6.5}{0.4}}
\def\sym{\Fund{6.5}{0.4} \kern-.5pt \Fund{6.5}{0.4}}
\Title{\vbox{\baselineskip12pt\hbox{hep-th/0509085}
\hbox{UCSD-PTH-05-14}
}}
{\vbox{\centerline{IR Free or Interacting? A Proposed Diagnostic}}}
\centerline{  Kenneth Intriligator}
\bigskip
\centerline{Department of Physics} \centerline{University of
California, San Diego} \centerline{La Jolla, CA 92093-0354, USA}
\medskip\centerline{\foot{Sabbatical address, Sept. '05 to June '06.}
School of Natural Sciences}\centerline{Institute for Advanced
Study}\centerline{Einstein Drive, Princeton NJ 08540}
\bigskip
\noindent
We present and discuss a conjectured criterion for determining whether a 4d quantum field theory  
is IR free, or flows to an interacting conformal field theory in the infrared: ``the correct
infrared phase is that with the larger conformal anomaly $a$".  A stronger conjecture is that ``an
operator can become IR free only if that results in a larger 
conformal anomaly $a$".    We test these conjectures in the context of $\N =1$ supersymmetric theories.  
They are verified to indeed predict the correct IR phase in every tested case, for a plethora 
of examples for which the infrared phase could already be determined on other grounds.  When applied to the still unsettled case of  $SU(2)$ with a chiral superfield in the isospin $3/2$ representation, the conjecture suggest that the IR phase is conformal rather than confining.

%\draftmode
\Date{September 2005}

\newsec{Introduction}

4d asymptotically free gauge theories have various possible IR phases.  In the
extreme IR, the massive degrees of freedom decouple, and remaining, massless
degrees of freedom are either IR free, or they're an interacting conformal field theory.  
As examples, pure Yang Mills confines, with no massless degrees of freedom
left in the IR.  $SU(N_c)$ QCD with $N_f$ massless quark flavors, provided that $N_f$ is sufficiently small, has confinement and $SU(N_f)_L\times SU(N_f)_R\rightarrow SU(N_f)_D$
chiral symmetry breaking; the theory in the extreme IR contains just the $N_f^2-1$
massless, Goldstone boson pions with interactions, given by the chiral Lagrangian, that are
all irrelevant in the extreme IR.  For $N_f$ sufficiently large, on the other hand, so that
the theory is just barely asymptotically free, the IR theory is an interacting conformal field theory
-- the Banks-Zaks RG fixed point -- corresponding to the zero  of the perturbative beta function at small coupling in this case \refs{\GW, \BZ}.  

%\GiesAS
\lref\GiesAS{
  H.~Gies and J.~Jaeckel,
  ``Chiral phase structure of QCD with many flavors,''
  arXiv:hep-ph/0507171.
  %%CITATION = HEP-PH 0507171;%%
}

%\NdiliNI
\lref\NdiliNI{
  F.~N.~Ndili,
  ``Phase transition critical flavor number of QCD,''
  arXiv:hep-ph/0508111.
  %%CITATION = HEP-PH 0508111;%%
}
These examples illustrate a general intuition.  When the massless matter content is closer to
(but still below) the asymptotic freedom bound, the IR phase is likely interacting conformal.
When the massless matter content is farther below the asymptotic freedom bound, the coupling runs to larger values, and the IR phase is then more likely to be an IR free theory of confined composites.    A long standing goal is to make this intuition more precise. Since we're unable to analytically solve the theory in the IR, the options are to either to go to
the lattice,  or to develop and employ  various diagnostics and strong coupling methods,
to try to gain insight into the IR phase.  This latter route (a.k.a. ``voodoo QCD")  
can be used to make predictions, but one can not
be absolutely certain as to whether or not they are correct.  An example of a strong coupling method are
the gap equations, which when applied to $SU(N_c)$ QCD with $N_f$ flavors suggest that the
theory has confinement and chiral symmetry breaking for all $N_f$ below a critical value $N^c_f$,
and is conformal for all $N_f$ in the range $N_f^c<N_f<11N_c/2$, with the critical value estimated to be $N_f^c\approx 4N_c$ \AppelquistRB.  See e.g. \refs{\GiesAS, \NdiliNI} for recent further analysis of this case. 

Confinement can occur with, or without chiral symmetry breaking, depending on the 
gauge group and matter content  (as exhibited by Seiberg
in the context of $\N =1$ supersymmetric SQCD \SeibergBZ).   't Hooft anomaly matching \thooftr\ strongly constrains the IR phase of theories with unbroken chiral symmetries: because the 't Hooft anomalies of global symmetries are unchanged along the RG flow, the IR free
spectrum of confined composites must contain massless fermions, that must match the 't Hooft
anomalies of the UV theory.   (When the chiral symmetry is broken, the Goldstone bosons instead saturate the 't Hooft
anomalies, via a Wess-Zumino term.)  A non-trivial, confining, solution of 't Hooft anomaly 
matching can be regarded as some kind of non-trivial evidence for the confining scenario.  
There are many examples of this in the context of supersymmetric gauge theories. One
particular, and still unsettled, example that we'll discuss further in what follows is $\N =1$
supersymmetric $SU(2)$ gauge theory with a matter chiral superfield in the $I=3/2$
representation \ISS.  But it was pointed out in \BCI\ that there are other examples, where highly
non-trivial 't Hooft anomaly matching, suggesting IR free confinement, can be a misleading fluke --  and the IR theory is instead interacting conformal. 

\nref\Zamo{A.~B.~Zamolodchikov,
``'Irreversibility' Of The Flux Of The Renormalization Group In A 2-D Field Theory,''
JETP Lett.\  {\bf 43}, 730 (1986)
[Pisma Zh.\ Eksp.\ Teor.\ Fiz.\  {\bf 43}, 565 (1986)].
%%CITATION = JTPLA,43,730;%%
}

\nref\Cardy{
  J.~L.~Cardy,
  ``Is There A C Theorem In Four-Dimensions?,''
  Phys.\ Lett.\ B {\bf 215}, 749 (1988).
  %%CITATION = PHLTA,B215,749;%%
}

\nref\AFGJ{D.~Anselmi, D.~Z.~Freedman, M.~T.~Grisaru and A.~A.~Johansen,
``Nonperturbative formulas for central functions of supersymmetric gauge
theories,''
Nucl.\ Phys.\ B {\bf 526}, 543 (1998)
[arXiv:hep-th/9708042].
%%CITATION = HEP-TH 9708042;%%
}

\nref\AEFJ{D.~Anselmi, J.~Erlich, D.~Z.~Freedman and A.~A.~Johansen,
``Positivity constraints on anomalies in supersymmetric gauge
theories,''
Phys.\ Rev.\ D {\bf 57}, 7570 (1998)
[arXiv:hep-th/9711035].
%%CITATION = HEP-TH 9711035;%%
}

\nref\IW{
K.~Intriligator and B.~Wecht,
``The exact superconformal R-symmetry maximizes a,''
Nucl.\ Phys.\ B {\bf 667}, 183 (2003)
[arXiv:hep-th/0304128].
%%CITATION = HEP-TH 0304128;%%
}

%\KutasovIY
\nref\KPS{
  D.~Kutasov, A.~Parnachev and D.~A.~Sahakyan,
  ``Central charges and U(1)R symmetries in N = 1 super Yang-Mills,''
  JHEP {\bf 0311}, 013 (2003)
  [arXiv:hep-th/0308071].
  %%CITATION = HEP-TH 0308071;%%
}

%\IntriligatorMI
\nref\twoadj{
  K.~Intriligator and B.~Wecht,
  ``RG fixed points and flows in SQCD with adjoints,''
  Nucl.\ Phys.\ B {\bf 677}, 223 (2004)
  [arXiv:hep-th/0309201].
  %%CITATION = HEP-TH 0309201;%%
}

%\KutasovUX
\nref\DKlm{
D.~Kutasov,
 ``New results on the 'a-theorem' in four dimensional supersymmetric field
theory,''
arXiv:hep-th/0312098.
%%CITATION = HEP-TH 0312098;%%
}

%\CsakiUJ
\nref\CsakiUJ{
  C.~Csaki, P.~Meade and J.~Terning,
  ``A mixed phase of SUSY gauge theories from a-maximization,''
  JHEP {\bf 0404}, 040 (2004)
  [arXiv:hep-th/0403062].
  %%CITATION = HEP-TH 0403062;%%
}
%\BarnesZN
\nref\BarnesZN{
  E.~Barnes, K.~Intriligator, B.~Wecht and J.~Wright,
  ``N = 1 RG flows, product groups, and a-maximization,''
  Nucl.\ Phys.\ B {\bf 716}, 33 (2005)
  [arXiv:hep-th/0502049].
  %%CITATION = HEP-TH 0502049;%%
}

%\OkudaME
\nref\Okuda{
  T.~Okuda and Y.~Ookouchi,
  ``Higgsing and Superpotential Deformations of ADE Superconformal Theories,''
  arXiv:hep-th/0508189.
  %%CITATION = HEP-TH 0508189;%%
}

Another diagnostic for the IR phase is the conjecture that the number of
massless degrees of freedom, suitably defined, decreases in RG flows to
the IR, in analogy with Zamolodchikov's c-theorem for 2d quantum field theories \Zamo.
We'll here be specifically interested in Cardy's conjectural a-theorem \Cardy,
that the  conformal anomaly 
coefficient $a$ (the coefficient of the Euler density term in $\ev{T_\mu ^\mu}$
when the theory is put on a fixed, curved, spacetime background)
satisfies $a_{UV}>a_{IR}$ for all 4d RG flows (and also 
$a_{IR} \geq 0$).   There is not yet
a definitive proof of Cardy's conjecture, but there is a huge amount of non-trivial
evidence for it, and not yet a single counter-example.  The evidence is especially
compelling in the context of supersymmetric gauge theories, where the conformal
anomaly $a$ can be computed exactly, even for interacting conformal field theories -- see
e.g. \refs{\AFGJ - \Okuda}.  

The successes of Cardy's conjecture are especially impressive
when one or both of the endpoints of the RG flow are interacting.  It's successful, but in
a less impressive way, when applied to RG flows with asymptotically free quarks and gluons
in the UV, and free confined composites in the IR.  E.g. for $SU(N_c)$ with $N_f$
quark flavors, the RG flow from asymptotically free quarks and gluons in the UV, 
to IR free pions in the IR, has 
\eqn\aqcdi{\eqalign{a_{UV}&=(N_c^2-1)a^{free}_1+2N_cN_fa^{free}_{1/2}\cr
a_{IR}^{confine}&=(N_f^2-1)a^{free}_0 
.}}
Here $a^{free}_j$ is the conformal anomaly coefficient $a$ for a free field of spin $j$;
which are found  (see e.g. in \DuffWM\ and references cited therein) to be given by
 $a^{free}_1=62 a^{free}_0$ for a massless spin 1 field and $a^{free}_{1/2}={11\over 2}a^{free}_0$ for a massless 2-component fermion.  For the case \aqcdi, $a_{UV}>a_{IR}$ is easily satisfied for
any asymptotically free $N_f$, so here Cardy's conjecture does not put any 
interesting constraint on the maximum $N_f$ for which confinement and chiral symmetry breaking
occurs.  Applied in this way, Cardy's conjecture does not help much in determining the IR
phase of the theory.

%\AnselmiFK
\lref\AnselmiFK{
  D.~Anselmi,
  ``Inequalities for trace anomalies, length of the RG flow, distance between
  the fixed points and irreversibility,''
  Class.\ Quant.\ Grav.\  {\bf 21}, 29 (2004)
  [arXiv:hep-th/0210124].
  %%CITATION = HEP-TH 0210124;%%
}

In this paper, we will propose and test a more conjectural diagnostic for determining the
IR phase of a given theory.  The idea is to compute the conformal anomaly coefficient 
$a$ in both the IR free phase scenario, and also in the
interacting conformal field theory phase scenario.  Our conjectural diagnostic is that {\it the correct IR phase is the one with the {\bf larger} value of the conformal anomaly $a$}.  
The motivation is that Cardy's conjecture suggests that $\Delta a\equiv a_{UV}-a_{IR}$ is
some measure of the RG distance between theories (see e.g. the discussion in 
\AnselmiFK).  So the conjecture that we'll explore here can be thought of as saying 
that the preferred IR phase is the one that's closest, in terms of RG distance, to the UV
starting point.  A perhaps reasonable objection to the conjecture is that the IR theory
is either interacting conformal or IR free  -- either it hits a zero of the beta functions before confinement sets in, or it does not -- so it is not clear whether or not there is any meaning in comparing them,
as if they were both viable possibilities.   In any case, we will show phenomenologically, by 
considering many understood examples, that the proposal correctly predicts the IR phase in every case.

%\AppelquistHR
\lref\AppelquistHR{
  T.~Appelquist, A.~G.~Cohen and M.~Schmaltz,
  ``A new constraint on strongly coupled field theories,''
  Phys.\ Rev.\ D {\bf 60}, 045003 (1999)
  [arXiv:hep-th/9901109].
  %%CITATION = HEP-TH 9901109;%%
}

Applied to $SU(N_c)$ QCD with $N_f$ massless quark flavors, the idea would be to compute
\eqn\aqcdii{\eqalign{a_{IR}^{confine}&\equiv a_{IR}^{free}=(N_f^2-1)a^{free}_0 \cr
a_{IR}^{interacting}&=(N_c^2-1)a^{free}_1+2N_cN_fa^{free}_{1/2}-(\hbox{interaction 
contributions}).}}
The interaction contributions in the conformal phase come from the presumed non-zero RG fixed point value of the gauge coupling, $g_*$.  As indicated in \aqcdii, the interactions are expected, from Cardy's conjecture, to always decrease the conformal anomaly $a$ -- this can
be verified perturbatively for Banks-Zaks RG fixed points.  In the Banks-Zaks limit of large,
barely asympotically free $N_f$, the interaction contributions in \aqcdii\ are $O(g_*^2)\ll 1$, and then comparing the two alternatives in \aqcdii\ we find $a_{IR}^{interact}>a_{IR}^{free}$, so our diagnostic correctly predicts the interacting conformal phase
in this limit.  (This is a trivial success: the Banks-Zaks fixed point has a small value of $\Delta a
\equiv a_{UV}-a_{IR}$ by assumption, and $a_{IR}^{free}$ couldn't have exceeded $a_{IR}^{interacting}$ in this limit without violating Cardy's conjecture.)   Now consider reducing
$N_f$: this increases the RG fixed 
point coupling $g_*$, and by Cardy's conjecture we expect that the interaction contributions
to $a_{IR}^{interacting}$ become more and more important, and always tend to further reduce
the value of $a_{IR}^{interacting}$.  It's possible, then, that there is a critical value $N_f^c$ where
the interaction effects are sufficiently strong so that $a_{IR}^{interacting}$ drops below the value
$a_{IR}^{free}$ in \aqcdii; our proposed diagnostic would then predict that the theory is 
IR free for this value of $N_f$ and below.   Unfortunately, we are unable at present to 
explicitly and reliably compute the interaction contributions to $a_{IR}^{interacting}$ in the strong coupling regime for this and other non-supersymmetric theories,  to compare the value of $N_f^c$ obtained in this way with those via diagnostics \refs{\AppelquistRB - \NdiliNI, \AppelquistHR}

In what follows, we will test and apply the proposed diagnostic in the context of 4d $\N =1$
supersymmetric gauge theories, where
 it's possible to exactly compute both $a_{IR}^{free}$ {\it and} $a_{IR}^{interacting}$, even at strongly interacting RG fixed points.  This will be reviewed in the following section. We will also discuss a stronger version
of our conjecture: ``operators can only become IR free if that increases $a$". 
 In section 3, we will test and apply the proposal in many examples.  As we will note there,
for the unsettled example \ISS\ of $SU(2)$ with matter in the $I=3/2$, the proposal predicts
that the IR theory is an interacting RG fixed point, rather than the confining phase.  But since
the proposal here is only a conjecture, this of course does not yet completely settle the issue of the IR phase of that example.  

\newsec{Exact results for $a$ in supersymmetric gauge theories}

4d $\N =1$ superconformal field theories have a conserved $U(1)_R$ symmetry, 
that's related by supersymmetry to the dilatation current.   We'll denote the charges of the 
fields under this superconformal R-symmetry by $R_*$.   It was shown in \AFGJ\ that the 
conformal anomaly $a_{SCFT}$ is related by supersymmetry to the 't Hooft anomalies
of this R-symmetry:
\eqn\afgja{a_{SCFT}={3\over 32}(3\Tr R_*^3-\Tr R_*).}
This formula is very powerful: using the power of 't Hooft anomaly matching, it allows $a_{SCFT}$ to be computed, even for strongly interacting IR fixed points, from the $R_*$ charges of the 
weakly coupled, asymptotically free, UV spectrum.  

Moreover, in \IW\ it was shown that the 
superconformal R-symmetry $R_*$ can be exactly determined by ``a-maximization": it's 
the local maximum of the  function 
\eqn\atrial{a_{trial}(R)={3\over 32}(3Tr R^3-\Tr R)}
over all possible, conserved R-symmetries.   The proposed diagnostic of the present paper fits
with the intuition of a-maximization: the conformal anomaly $a$ wants to be as big as possible.
This helps ensure that  $a$ then decreases in RG flows to the IR \IW.

As will be important in what follows, theories can develop accidental symmetries in the IR.
To be concrete, suppose that a set of composite, gauge invariant operators -- let's call
them $X$ -- decouple in the extreme IR from the other operators, i.e. 
correlation functions involving $X$ can be computed by treating $X$ as decoupled, free fields.   The 
effect of this on $a_{SCFT}$, and on the a-maximization procedure, can be included via 't Hooft anomaly matching \refs{\AEFJ,\KPS}.  The idea is that the contribution of the
fields $X$  to the 't Hooft anomalies in \afgja\ can be isolated, using anomaly matching,  as:
\eqn\axis{a_X(R_X)={3\over 32}\hbox{dim}(X)\left[3(R_X-1)^3-(R_X-1)\right],}
where $\hbox{dim}(X)$ is the number of operators $X$ and $R_X$ is their superconformal
R-charge.  Without accidental symmetries, $R_X$ would be fixed to $R_X=R^{(0)}(X)$, the
naive R-charge of the composite operator, obtained by adding the R-charges of the fields in it. 
But when $X$ is a free field, there is an additional $U(1)_X$ accidental symmetry, and a-maximization requires maximization over the full space of possible R-symmetries, including
mixing with $U(1)_X$.  This means that $R_X$ becomes an additional variable to
locally maximize over, without being constrained to equal $R^{(0)}(X)$.  Maximizing \axis\ 
over $R_X$ gives $R_X=2/3$, the correct value for a free field.  The upshot of this change is to
replace \afgja\ or \atrial\ with:
\eqn\aaccident{\eqalign{a\rightarrow a&+a_X(2/3)-a_X(R^{(0)}(X))\cr
=a&+{1\over 96}\hbox{dim}(X)(2-3R^{(0)}(X))^2(5-3R^{(0)}(X)).}}

There is a unitarity bound $R(X)\geq 2/3$ for gauge invariant, spinless chiral primary operators
$X$ (since $\Delta X\geq 1$), with equality if and only if $X$ is a free field.   If some operator $X$ would otherwise appear to violate this bound, the picture is that, in RG flow to the IR fixed point,
$R(X)$ flows down until it hits $2/3$, and thereafter $X$ remains free, with $R(X)=2/3$. 
  In this case, because $R^{(0)}(X)<2/3$, the modification of \aaccident\ increases the value of $a$.  This is reasonable:
the correction in \aaccident\ comes from maximizing the function $a_{trial}(R)$ over a bigger
space of possible trial R-symmetries, 
and  it's reasonable that maximizing $a_{trial}$ over the larger space of possibilities leads to
a bigger maximal value.

But operators with $R^{(0)}(X)>2/3$ can also turn out to be free fields.  There can be an accidental $U(1)_X$ symmetry, which leads to the same modification \aaccident, even
when it is not required by the unitarity bound.  In particular, in a theory that's
actually confining in the IR, the confining composite operators are all IR free fields, regardless 
of their value of $R^{(0)}$.  We will illustrate this with 
many examples in the following section.   In this case, we see immediately from \aaccident\ 
that the $U(1)_X$ accidental symmetry leads to an increase in $a_{CFT}$ if, and only if, 
$R^{(0)}(X)<5/3$.  If $R^{(0)}(X)>5/3$, we have a
peculiar situation: locally maximizing $a_{trial}$ over the bigger space of possible R-symmetries, with $U(1)_X$ included,  then leads to a smaller value at the maximum.  This is 
simply because, when  $R^{(0)}(X)>5/3$, locally maximizing the cubic function \axis\
over $R_X$ will lead to a smaller value of $a$ than if $R_X$ were just fixed to $R^{(0)}(X)$.

Though the math is clear, having $R^{(0}(X)>5/3$ operators become free fields runs counter to our intuition that the conformal anomaly $a_{CFT}$ always wants to be as big as
possible.  This motivates the {\bf conjecture}: {\it operators only become IR free if that 
leads to a larger value of $a$, i.e. operators with $R^{(0)}(X)>5/3$ do not become IR free.}
 All of the understood examples, to be discussed in the following section, are compatible with this conjecture.  This conjecture is a
stronger version of the one in the introduction, since it states that {\it every} IR free operator
should contribute positively to $a$, while the weaker conjecture is that their sum must
be positive.  The conjecture here can also apply to IR free operators in a theory that
remains partially interacting.

\newsec{Examples}

\subsec{$\N =1$ $SU(N_c)$ SQCD, with $N_f$ massless fundamental flavors}

We will here be interested in the vacuum at the origin of the moduli space of vacua for
$N_f>N_c$ \refs{\SeibergBZ, \NSd}, where the global symmetries are all unbroken.  In particular, there is a conserved $U(1)_R$ symmetry, with anomaly free charges
\eqn\sqcdr{R_{cons}(Q)=R_{cons}(\widetilde Q)={N_f-N_c\over N_f}.}  
The gauge invariant composite operators include the ``mesons" $M=Q\widetilde Q$ 
and the ``baryons" $B=Q^{N_c}$ and $\widetilde B=\widetilde Q ^{N_c}$
(suppressing flavor and color indices), with 
$U(1)_R$ charges 
\eqn\sqcdrr{R^{(0)}(M)=2\left({N_f-N_c\over N_f}\right), \qquad R^{(0)}(B)=R^{(0)}(\widetilde B)=N_c\left({N_f-N_c\over N_f}\right).}
The superscript is a reminder that these might be
corrected by accidental symmetries. 

Consider first $N_f=N_c+1$ massless flavors. The vacuum at the origin has confinement
without chiral symmetry breaking, with the massless spectrum consisting of \SeibergBZ\ 
 $N_f^2$ mesons $M$, $N_f$ baryons $B$, and $N_f$ anti-baryons $\widetilde B$.
In the extreme IR, these confined operators are all free fields.
Let us clarify this: the IR theory of \SeibergBZ\ contained also a superpotential term for these
fields,  $W_{conf}\sim MB\widetilde B-\det M$, but
any superpotential term for gauge neutral fields (other than quadratic mass terms) is always irrelevant in the extreme IR.   So the superpotential $W_{conf}$ is irrelevant\foot{More precisely, $W_{conf}$ is ``dangerously irrelevant," in that it becomes relevant upon
deforming the theory, by giving the flavors masses or moving away from the origin of the
moduli space of vacua.  But we're only considering the undeformed theory above, so this
distinction is immaterial.}
in the extreme IR, $W_{conf}\rightarrow 0$, and the IR theory 
consists of just the confined, free massless fields $M_{i\tilde j}$, $B^i$, $\widetilde B^{\tilde i}$.  The IR limit of the RG flow thus has $R_*(M)=R_*(B)=R_*(\widetilde B)=2/3$, and thus, using
\afgja\ with this IR spectrum, 
\eqn\aconf{a_{IR}^{free}={3\over 32}\left(3(R_*-1)^3-(R_*-1)\right)(N_f^2+N_f+N_f)={1\over 48}(N_f^2+2N_f).}

Evaluating the 't Hooft anomalies in \afgja\ using the UV asymptotically free spectrum of quarks
and gluons, for general $N_c$, and $N_f$, with the conserved R-charges \sqcdr\ yields:
\eqn\aois{\eqalign{a^{(0)}&={3\over 32}\left[2(N_c^2-1)+2N_cN_f\left(3(-{N_c\over N_f})^3-(-{N_c\over N_f})\right)\right],\cr
&={3\over 32}\left[4N_c^2-2-6{N_c^4\over N_f^2}\right],}}
where the superscript indicates that we have not yet accounted for the accidental symmetries.
The same result would be obtained in a dual description, since the dual 't Hooft anomalies
should match.  In particular, for $N_f=N_c+1$, the 't Hooft
anomalies match between the UV and confined IR free spectrum \SeibergBZ, so \aois\ is also obtained  upon evaluating the 't Hooft anomalies in \afgja\ using the confined
IR free spectrum of mesons and baryons, with R-charges \sqcdrr. The result \aconf\ is then obtained upon accounting for the accidental symmetries, using the formula \aaccident\ 
for the IR free operators $M$, $B$, $\widetilde B$:
\eqn\sqcdacone{a_{IR}^{free}=a^{(0)}+{N_f^2\over 96}(2-3R^{(0)}(M))^2(5-3R^{(0)}(M))+
{2N_f\over 96}(2-3R^{(0)}(B))^2(5-3R^{(0)}(B)).}

Because $R^{(0)}(M)=2/N_f<2/3$, the unitarity bound shows that the operators 
$M$ {\it had} to become free fields.  On the other hand, the operators $B$ and $\widetilde B$ 
have $R^{(0)}(B)=R^{(0)}(\widetilde B)=(N_f-1)/N_f$, which is safely above the unitarity bound
for general $N_f$.  So there is an alternative (incorrect) scenario to the above, where rather
than having all operators confine to free fields, the theory flows to an interacting conformal 
field theory in the IR, where the operators $M$ are free, but the operators $B$ and $\widetilde B$ are not free.    In this hypothetical scenario, the value
of $a_{IR}$ for the IR theory is given by \aois, corrected only by the accidental symmetry \aaccident\  for the meson fields $M$:
\eqn\sqcdaconl{a_{IR}^{interacting}=a^{(0)}+{N_f^2\over 96}(2-3R^{(0)}(M))^2(5-3R^{(0)}(M)).}

Because $R^{(0)}(B)=(N_f-1)/N_f<5/3$, we can immediately see that the additional
baryon contribution in $a_{IR}^{free}$ \sqcdacone, as compared with $a_{IR}^{interacting}$
\sqcdaconl, is positive, so
\eqn\sqcdcomp{a_{IR}^{free}>a_{IR}^{interacting}.}
Our proposed diagnostic then predicts that the actual IR phase is the IR free confining scenario, rather than the conformal one; this prediction is indeed the correct answer \refs{\SeibergBZ, \NSd}.

Now let us consider general numbers of flavors, in the range $N_c+1\leq N_f<3N_c$, with the
upper limit for asymptotic freedom.  We again consider the possibility that the IR theory is an
interacting conformal field theory, with the only free fields given by those gauge invariant operators
which hit or violate the unitarity bound, $R(X)\geq 2/3$.  Again using \sqcdr\ for the
superconformal R-charges, this gives the result 
\eqn\sqcdaconfx{\eqalign{a^{interacting}_{IR}&=a^{(0)}={3\over 32}\left[4N_c^2-2-6{N_c^4\over N_f^2}\right] \qquad\hbox{for}\quad {3\over 2}N_c<N_f<3N_c, \cr
a^{interacting}_{IR}&=a^{(0)}+{1\over 96}N_f^2(2-3R^{(0)}(M))^2(5-3R^{(0)}(M))\qquad\hbox{for}
\quad N_c+1\leq N_f\leq {3\over 2}N_c,}}
where $R^{(0)}(M)=2(N_f-N_c)/N_f$.  For $N_f\leq 3N_c/2$, $R^{(0)}(M)\leq 2/3$, below 
the unitarity bound, so then $M$ must be free, with the resulting additional contribution in 
\sqcdaconfx.  

For $N_f>{3\over 2}N_c$, the interacting conformal scenario is the only
known viable possibility.  
But for $N_f\leq {3\over 2}N_c$ there is another IR phase scenario, which was argued in \NSd\ to be the correct answer: the $SU(N_f-N_c)$ gauge
fields and quarks of the non-asymptotically free, and hence  IR free, magnetic dual.  See
e.g. \ISrev\ for some discussion of the characteristics of the free magnetic phase, e.g. the 
potential for sources.   Here we will just use the fact that it's free in the extreme IR.  
  The conformal anomaly of the free magnetic phase is 
\eqn\sqcdfma{\eqalign{a^{free}_{IR}&=a^{(0)}+{1\over 96}N_f^2(2-3R^{(0)}(M))^2(5-3R^{(0)}(M))\cr &+ {1\over 96}2N_f(N_c-N_f)(2-3R^{(0)}(q))^2(5-3R^{(0)}(q)) \qquad\hbox{for}\quad N_c+1<N_f<{3\over 2}N_c.}}
Again, $a^{(0)}$ is given by \sqcdaconfx, which is also obtained if the 't Hooft anomalies
in  \afgja\ are evaluated in the magnetic dual description \NSd, with 
its conserved R-charges
\eqn\sqcdmr{R^{(0)}(M)=2\left({N_f-N_c\over N_f}\right), \qquad R^{(0)}(q)=R^{(0)}(\widetilde q)={N_c\over N_f},}  
by the 't Hooft anomaly matching for the dual theories \NSd.  
 The additional terms in \sqcdfma\ are to account, as in \aaccident, for the 
fact that the field $M$, $q$, and $\widetilde q$ are all IR free fields in the free magnetic scenario.  
Comparing \sqcdfma\ with \sqcdaconfx, the only difference is 
in the last line of \sqcdfma, coming from $q$ and $\widetilde q$ being IR free in
the free magnetic dual scenario.    Because $R^{(0)}(q)=N_c/N_f<5/3$, this additional contribution is positive,
so $a_{IR}^{free}>a_{IR}^{interacting}$.    Our
proposed diagnostic then predicts that free magnetic scenario is preferred over the
(partially) interacting conformal scenario for $N_f<{3\over 2}N_c$.  Based on the many other duality checks in \NSd, this is believed to be the correct answer.

%\IntriligatorID
\lref\IntriligatorID{
  K.~A.~Intriligator and N.~Seiberg,
  ``Duality, monopoles, dyons, confinement and oblique confinement in
  supersymmetric SO(N(c)) gauge theories,''
  Nucl.\ Phys.\ B {\bf 444}, 125 (1995)
  [arXiv:hep-th/9503179].
  %%CITATION = HEP-TH 9503179;%%
}
%\IntriligatorNE
\lref\IntriligatorNE{
  K.~A.~Intriligator and P.~Pouliot,
  ``Exact superpotentials, quantum vacua and duality in supersymmetric SP(N(c))
  gauge theories,''
  Phys.\ Lett.\ B {\bf 353}, 471 (1995)
  [arXiv:hep-th/9505006].
  %%CITATION = HEP-TH 9505006;%%
}
It is very similarly verified for the dualities for $SO$ and $Sp$ with fundamentals \refs{\IntriligatorID, \IntriligatorNE} that, in the free magnetic phase, the free magnetic
quarks indeed satisfy $R^{(0)}(q)<5/3$.  

\subsec{Other examples with IR free ``confining" phases}

Our conjectures can be similarly checked in the multitude of other supersymmetric
examples which have been argued to have an IR free phase.   The process is to
compute the superconformal R-charges $R^{(0)}(X)$ of the basic gauge invariant chiral operators in a hypothetical, interacting conformal phase.  Those operators with $R^{(0)}(X)$
at or below the unitarity bound are forced to be IR free in any case.  
 Any basic operators with $R^{(0)}(X)>2/3$ are more interesting,
since they could have been interacting.  Our conjecture is that, if they're actually IR free,
these basic operators  would have had $R^{(0)}(X)<5/3$ in the hypothetical interacting
phase.  Then having them be IR free, rather than interacting, leads to larger $a$, 
and so  $a_{IR}^{free}>a_{IR}^{interacting}$.

%\DottiWN
\lref\DottiWN{
  G.~Dotti and A.~V.~Manohar,
  ``Supersymmetric gauge theories with an affine quantum moduli space,''
  Phys.\ Rev.\ Lett.\  {\bf 80}, 2758 (1998)
  [arXiv:hep-th/9712010].
  %%CITATION = HEP-TH 9712010;%%
}
A set of examples, with t Hooft anomaly matchings suggesting an IR free
spectrum,  was presented in \DottiWN.   The examples
with $\mu >\mu _{adj}$ ($\mu$ is the index of the matter representation and $\mu _{adj}$ that of the adjoint) will be discussed in the following subsection.  The $\mu <\mu _{adj}$ 
examples were  referred to as T1-T6 in  \DottiWN, where e.g. T6 is $SO(14)$ with a single matter field in the spinor representation.  Because $\mu <\mu _{adj}$, the matter fields in these examples 
all have $R=-(\mu _{adj}-\mu )/\mu <0$, so all gauge invariant 
composite operators also have $R^{(0)}(X)<0$. The unitarity bound then requires them
to all be IR free fields.  So the theories T1-T6 necessarily confine to IR free fields (on
their $W_{dyn}=0$ branch) -- there is no viable interacting conformal scenario.  

%\PouliotME
\nref\PouliotME{
  P.~Pouliot,
  %``Duality in SUSY $SU(N)$ with an Antisymmetric Tensor,''
  Phys.\ Lett.\ B {\bf 367}, 151 (1996)
  [arXiv:hep-th/9510148].
  %%CITATION = HEP-TH 9510148;%%
}
%\PouliotSK
\nref\PouliotSK{
  P.~Pouliot and M.~J.~Strassler,
  ``A Chiral $SU(N)$ Gauge Theory and its Non-Chiral $Spin(8)$ Dual,''
  Phys.\ Lett.\ B {\bf 370}, 76 (1996)
  [arXiv:hep-th/9510228].
  %%CITATION = HEP-TH 9510228;%%
}
%\PouliotZH
\nref\PouliotZH{
  P.~Pouliot and M.~J.~Strassler,
  ``Duality and Dynamical Supersymmetry Breaking in $Spin(10)$ with a Spinor,''
  Phys.\ Lett.\ B {\bf 375}, 175 (1996)
  [arXiv:hep-th/9602031].
  %%CITATION = HEP-TH 9602031;%%
}

%\KawanoBD
\nref\KawanoBD{
  T.~Kawano,
  ``Duality of N=1 Supersymmetric SO(10) Gauge Theory with Matter in the
  Spinorial Representation,''
  Prog.\ Theor.\ Phys.\  {\bf 95}, 963 (1996)
  [arXiv:hep-th/9602035].
  %%CITATION = HEP-TH 9602035;%%
}

%\ChoBI
\nref\ChoBI{
  P.~L.~Cho and P.~Kraus,
  ``Symplectic SUSY gauge theories with antisymmetric matter,''
  Phys.\ Rev.\ D {\bf 54}, 7640 (1996)
  [arXiv:hep-th/9607200].
  %%CITATION = HEP-TH 9607200;%%
}
%\CsakiZB
\nref\CsakiZB{
  C.~Csaki, M.~Schmaltz and W.~Skiba,
  ``Confinement in N = 1 SUSY gauge theories and model building tools,''
  Phys.\ Rev.\ D {\bf 55}, 7840 (1997)
  [arXiv:hep-th/9612207].
  %%CITATION = HEP-TH 9612207;%%
}
%\BerkoozBB
\nref\BerkoozBB{
  M.~Berkooz, P.~L.~Cho, P.~Kraus and M.~J.~Strassler,
  ``Dual descriptions of SO(10) SUSY gauge theories with arbitrary numbers  of
  spinors and vectors,''
  Phys.\ Rev.\ D {\bf 56}, 7166 (1997)
  [arXiv:hep-th/9705003].
  %%CITATION = HEP-TH 9705003;%%
}

The ``s-confining" theories of e.g. \refs{\PouliotME - \BerkoozBB}, provide many more
examples.  Given their many checks, we believe that these examples are indeed
IR free, rather than interacting.  Our conjecture 
can then be tested,  by verifying that all of the IR free operators of every example indeed 
would have had $R^{(0)}(X)<5/3$ in any viable interacting phase.   Scanning the
 examples, this can be immediately verified in those cases where the superconformal R-symmetry is uniquely determined on symmetry grounds.  For those examples with matter in different representations 
 of the gauge group, on the other hand, a-maximization is needed to determine the superconformal R-charges $R^{(0)}(X)$ in the hypothetical conformal phase.  This is straightforward to implement, but
it would take too long here to check every possible example.   Let us illustrate one example.

Consider $SU(N)$ gauge theory with matter $A$ in the two-index anti-symmetric tensor representation $\asym$, $\overline A$ in the conjugate representation, and $N_f$ flavors $Q_f$ and $\overline  Q_f$ of fundamental and anti-fundamental matter.  The confining
case of \CsakiZB\ is $N_f=3$, but we'll temporarily keep $N_f$ as a free parameter. 
The superconformal R-charges of the fields are constrained by the anomaly free condition
to satisfy $R_A=(N_f-2-N_fy)/(N-2)$, with $y\equiv R_Q$.  
The trial a-function, before accounting for accidental symmetries, is
\eqn\antiat{a^{(0)}_t(y)={3\over 32}\left[2(N-1)+N(N-1)(3R_A^3-9R_A^2+8R_A)
+2NN_f(3(y-1)^3-(y-1))\right].}
As an interesting limit, let us consider the situation for $N\gg 1$,
and simply the formulae by keeping only leading order in large $N$.  So $R_A\approx 
(N_f-2-N_fy)/N$ and \antiat\ becomes
\eqn\antiato{a_t^{(0)}\approx {3N\over 16}\left[5N_f(1-y)-7+3N_f(y-1)^3
\right].}

If we were to maximize this, setting $N_f=3$, we'd obtain $y^{(0)}=1-\sqrt{15/27}\approx 
0.255$.  But we must correct, as in \aaccident, for the many gauge invariant operators
which are at or below the unitarity bound, and hence actually free.   The gauge invariant operators are listed in case 3.1.4
or 3.1.5 of \CsakiZB\ for odd or even N, respectively; because $N$ is large, there is no significant difference here between the even and odd $N$ cases.  To be comparable with \antiato, which is order $N$,
the only significant accidental symmetry contributions are those coming from summing a
series of operator contributions, i.e. the operators $M_j=Q(A\overline A)^{j-1}\overline Q$, $H_j=\overline A(A\overline A)^{j-1} Q^2$, $\overline H_j=A(A\overline A)^{j-1}\overline Q^2$, and $T_j=(A\overline A)^j$, for $j=1\dots N/2$.  The sum of terms \aaccident\ can then be approximated by an 
integral, as in \refs{\KPS ,\BarnesZN}
\eqn\accsum{{1\over 96}\sum _{j=1}^p[2-3R_j]^2[5-3R_j]\approx {1\over 288\beta}\int _{2-3R_p}^{2-3\alpha}u^2(3+u)du,}
where $\alpha$ and $\beta$ are defined by $R_j\equiv \alpha +(j-1)\beta$ (and $u\equiv 
2-3R_j$), and $p$ is either solved for by setting $R_{p}=2/3$, or if all the operators are
below the unitarity bound, so $R_p=2/3$ has no solution, 
then $p=j_{max}=N/2$.  Applying this for the 
 operators $M_j$, $H_j$, $\overline H_j$ and $T_j$, with $N_f=3$, modifies 
\antiato\ to 
\eqn\antiat{a_t(y)\approx a_t^{(0)}(y)+{15\over 48}N(1-3y)^2(1-y)+{1\over 768}N(17+27y+243y^2+729y^3).}
Maximizing this w.r.t. $y$ gives $y_*=1/3$, which implies that {\it every} operator is at, or
below the unitarity bound.  So this 
particular class of examples actually has no viable alternative, interacting conformal field theory scenario:  the IR theory is necessarily free.  

\lref\DKNSAS{D.~Kutasov, A.~Schwimmer and N.~Seiberg,
``Chiral Rings, Singularity Theory and Electric-Magnetic Duality,''
Nucl.\ Phys.\ B {\bf 459}, 455 (1996)
[arXiv:hep-th/9510222].
%%CITATION = HEP-TH 9510222;%%
}
%\CsakiFM
\lref\CsakiFM{
  C.~Csaki and H.~Murayama,
  ``New confining N = 1 supersymmetric gauge theories,''
  Phys.\ Rev.\ D {\bf 59}, 065001 (1999)
  [arXiv:hep-th/9810014].
  %%CITATION = HEP-TH 9810014;%%
}

\lref\DKi{
D.~Kutasov,
``A Comment on duality in N=1 supersymmetric nonAbelian gauge
theories,''
Phys.\ Lett.\ B {\bf 351}, 230 (1995)
[arXiv:hep-th/9503086].
%%CITATION = HEP-TH 9503086;%%
}

\lref\DKAS{
D.~Kutasov and A.~Schwimmer,
``On duality in supersymmetric Yang-Mills theory,''
Phys.\ Lett.\ B {\bf 354}, 315 (1995)
[arXiv:hep-th/9505004].
%%CITATION = HEP-TH 9505004;%%
}
%\KleinUC
\lref\KleinUC{
  M.~Klein,
  ``More confining N = 1 SUSY gauge theories from non-Abelian duality,''
  Nucl.\ Phys.\ B {\bf 553}, 155 (1999)
  [arXiv:hep-th/9812155].
  %%CITATION = HEP-TH 9812155;%%
}

Yet another class of examples of theories is 
$SU(N)$ gauge theory, with an adjoint $X$, $F$ fundamentals and anti-fundamentals $Q$
and $\overline Q$, and superpotential
\eqn\waki{W_{tree}=h\Tr X^{k+1},}
which has been argued to confine \refs{\DKNSAS , \CsakiFM} when $N=kF-1$, using
the duality of \refs{\DKi, \DKAS}.  We know from the a-maximization analysis \KPS\ 
of the interacting SCFT with $W_{tree}=0$ that the superpotential \waki\ 
is relevant, for all $k$, given the above relation between the colors and flavors.  
Thanks to the superpotential \waki, we do not need to use a-maximization to determine
the superconformal R-charges of a hypothetical conformal phase, we simply have 
$R(X)=2/(k+1)$ and $R(Q)=R(\overline Q)=1-2(kF-1)/(k+1)F$.    Using these, we
compute the R-charges $R^{(0)}(X)$ of the confined spectrum of operators. 
The generalized mesons $M_i=(\overline Q X^{i-1}Q)$, with $i=1\dots k$, are
below the unitarity bound, and hence IR free in any case.  The generalized
baryons,  $B=(Q^F(XQ)^F\dots (X^{k-1}Q)^{F-1})$, and $\overline B=(\overline Q^F(X\overline Q)^F\dots (X^{k-1}\overline Q)^{F-1})$, on the other hand, have $R^{(0)}(B)=1-(2/F(k+1))$, 
which is generally above the unitarity bound, and thus there is an alternative
scenario where they're interacting.  Because $R^{(0)}(B)<5/3$, our diagnostic correctly 
predicts they're actually IR free.    Note that the operators $T_j=\Tr X^{j}$ are not included,
because they are not IR free massless fields --  they pair up with quadratic 
superpotential mass terms.  

%\IntriligatorAX
\lref\ILS{
K.~A.~Intriligator, R.~G.~Leigh and M.~J.~Strassler,
 ``New examples of duality in chiral and nonchiral supersymmetric gauge
theories,''
Nucl.\ Phys.\ B {\bf 456}, 567 (1995)
[arXiv:hep-th/9506148].
%%CITATION = HEP-TH 9506148;%%
}

%\IntriligatorFF
\lref\kispso{
K.~A.~Intriligator,
``New RG fixed points and duality in supersymmetric SP(N(c)) and SO(N(c)) gauge
theories,''
Nucl.\ Phys.\ B {\bf 448}, 187 (1995)
[arXiv:hep-th/9505051].
%%CITATION = HEP-TH 9505051;%%
}

%\LeighQP
\lref\LeighQP{
  R.~G.~Leigh and M.~J.~Strassler,
  ``Duality of Sp(2N(c)) and S0(N(c)) supersymmetric gauge theories with
  adjoint matter,''
  Phys.\ Lett.\ B {\bf 356}, 492 (1995)
  [arXiv:hep-th/9505088].
  %%CITATION = HEP-TH 9505088;%%
}

There are many generalizations of this example, discussed in \refs{\CsakiFM, \KleinUC},
which can be argued to confine using the dualities of \refs{\kispso, \LeighQP, \ILS}.  Our
conjecture that all IR free massless operators would have had 
$R^{(0)}(X)<5/3$ in a hypothetical interacting conformal scenario can be checked in all of these examples.  As in the above example, there are operators $T_j$ which pair up to get masses, via quadratic superpotential terms. Our conjecture
that  $R^{(0)}(X)<5/3$ does not apply to them, because they're not
IR free massless fields.

\subsec{Misleading anomaly matching examples}

In \BCI\ it was pointed out that highly non-trivial anomaly matching, suggesting confined IR
free fields, can sometimes be a misleading fluke.  The example of \BCI\ was $SO(N)$ with a matter
field $S$ in the $\sym$.  
The conserved, anomaly free R-charge is $R(S)=4/(N+2)$, and the $\Tr U(1)_R$ and $\Tr U(1)_R^3$ 't Hooft anomalies of the asymptotically free UV fields were shown to match with those 
of the composite operators ${\cal O}_n=\Tr S^n$, $n=2\dots N$, suggesting that the IR theory
confines to a free field theory of these composites.  But by deforming the theory, both by moving away from the origin of the moduli space of vacua, and by adding superpotential terms, it was
shown in \BCI\ that the non-trivial anomaly matching was misleading: the theory at the origin can be shown to instead be an interacting conformal field theory.  Here we note that our conjectured condition for accidental symmetries and our diagnostic for
confinement, applied to this case, is again successful.   

The operators ${\cal O}_n$ have R-charge $R_n^{(0)}=R({\cal O}_n)=4n/(N+2)$, for $n=2\dots N$.  Note that all $n\leq (N+2)/6$ have $R_n^{(0)}\leq 2/3$, naively below
the unitarity bound, so these operators must become free fields in any case.  The
issue is what happens with the operators for $n>(N+2)/6$.  In the (incorrect)
``confining" scenario, the ${\cal O}_n$ are IR free fields for all $n=2\dots N$.  In
the ``conformal" scenario, at least some of the operators with $n>(N+2)/6$ remain interacting -- to be concrete, we'll suppose that all the operators with $n>(N+2)/6$ remain interacting. 
Note that for all  $n>5(N+2)/12$, the operators ${\cal O}_n$ have $R^{(0)}(X)>5/3$, above
our conjectural upper bound for an operator to
become a free field.  Let us now compare the conformal anomaly in the two scenarios.
Without accounting for any accidental symmetries, the conformal anomaly obtained from \afgja\ would be
\eqn\aoson{a^{(0)}={3\over 32}\left[3{(N-1)(5N^2-4N+4)\over (N+2)^2}-(N-1)\right].}
Now we need to add the accidental symmetry corrections \aaccident, to obtain
\eqn\asonx{\eqalign{a_{IR}^{interacting}&=a^{(0)}+\sum _{n=2}^{(N+2)/6}{1\over 96}(2-3R_n^{(0)})^2(5-3R^{(0)}_n)\approx a^{(0)}+{N\over 96}\cr
a_{IR}^{free}&=a^{(0)}+\sum _{n=2}^{N}{1\over 96}(2-3R_n^{(0)})^2(5-3R^{(0)}_n)={1\over 48}(N-1).}}
In the last expression for $a_{IR}^{interacting}$ we simplified by considering $N\gg 1$, using \accsum, and dropping terms  which are sub-leading in $1/N$.  
The last expression for $a_{IR}^{free}$ can also be obtained immediately by applying \afgja\ directly to the spectrum of operators ${\cal O}_n$, with the free field R-charges $R({\cal O}_n)=2/3$.   Comparing the expressions in \asonx, we here find 
\eqn\confp{a_{IR}^{interacting}>a_{IR}^{free},}
for all $N$; e.g. for $N\gg 1$, we have $a_{IR}^{interacting}\approx 127N/96$ and $a_{IR}^{free}\approx 2N/96$.   So our proposed diagnostic will predict that the IR conformal scenario is favored over the
IR confining scenario,  in agreement with what was argued to be the correct answer in 
\BCI.

The above example was referred to as $T10$ in the classification table of \DottiWN.   The theory  T7 in the classification of \DottiWN\ is the theory of \ISS, which will be separately discussed
in the following subsection.  All of the other examples of \DottiWN\ with $\mu >\mu _{adj}$, 
i.e. their T8-T11, can also be argued to have misleading anomaly matchings, much 
as in \BCI.  These other theories are, T8: $SU(8)$ with matter in the 4-index antisymmetric tensor representation (the {\bf 70}); T9: $Sp(4)$ with matter in the 4-index
antisymmetric tensor (the {\bf 42}); T11: $SO(16)$ with matter in the spinor (the {\bf 128}).
As pointed out in \DottiWN, the dynamics of all of the theories T8-T11 can be related to
that of $SO(N)$ with 2-index symmetric tensor, so  the analysis of \BCI\ shows that all have interacting conformal, rather than confining phase, at the origin of the moduli space of vacua. 
For all of these theories, we find that $a_{IR}^{interacting}>a_{IR}^{free}$, so our diagnostic
passes the test of predicting the correct IR phase for all of these cases.  

Again, the reason why $a_{IR}^{free}$ is less than $a_{IR}^{interacting}$ is because 
the (incorrect) confining scenario entails accidental symmetries for  gauge invariant operators $X$ that would otherwise have $R^{(0)}(X)>5/3$.  E.g. for theory T11, $SO(16)$ with matter $S$ in the spinor, the  anomaly free R symmetry is $R(S)=1-(\mu /\mu _{adj})=1/8$ and the
basis of gauge invariants are (\DottiWN, table II)  $O_n=S^n$ for $n=$2, 8, 12, 14, 18, 20, 24, 30.  The $n=2$
operator must be free in any case, because of the unitarity bound.  The operators with $n=14$
and higher have $R^{(0)}>5/3$, and our stronger conjecture is that none of these can be IR free.

\subsec{$SU(2)$ with $Q$ in the $I=3/2$ representation}

This theory was discussed in \ISS.  The anomaly free $U(1)_R$ symmetry has $R(Q)=3/5$.
The basic gauge invariant composite that can be formed is $X=Q^4$, with $R^{(0)}(X)=12/5$.
The $\Tr R^3$ and $\Tr R$ 't Hooft anomalies of the UV fields happen to match
with those of $X$: $\Tr R=3+4(3/5-1)=(12/5-1)$ and $\Tr R^3=3+4(3/5-1)^3=(12/5-1)^3$.
This matching, which suggests that the IR theory at the origin of the moduli space of vacua 
consists of just the confined, IR free field $X$,
``seems too miraculous to be a coincidence".  But as noted in \ISS, there is also the possibility
that the theory at the origin is instead an interacting conformal field theory.   Let us discuss
the two possibilities:

In the interacting, conformal field theory scenario, the value of $a$ is, using  \afgja, 
\eqn\issal{a_{IR}^{interacting}={3\over 32}\left(3{343\over 125}-{7\over 5}\right)=0.6405}
(assuming that there are no overlooked accidental symmetries).
Now consider deforming the theory by the superpotential $W_{tree}=\lambda X$.  
Because $R^{(0)}(X)>2$, this superpotential is irrelevant in the IR, $W_{tree}\rightarrow 0$, 
and the deformed theory simply flows back again to the same interacting conformal field theory
in the IR -- adding $W_{tree}$ does nothing.

In the IR free, confining scenario, the value of $a$ is that of the massless free field $X$:
\eqn\issae{a_{IR}^{free}={3\over 32}\left(3({2\over 3}-1)^3-({2\over 3}-1)\right)={1\over 48}.}
In this case, deforming the theory by $W_{tree}=\lambda X$ is a relevant deformation,
and it would break supersymmetry \ISS, dynamically because of the confinement (classically,
there'd be a supersymmetric vacuum at $Q=0$).  This is the scenario that was hoped for in \ISS.  

The correct IR phase remains an unsettled question. The difficulty, as compared with other supersymmetric theories, is that there are so few ways to deform the 
theory.   

We note that
\eqn\aisscomp{a^{interacting}_{IR}>a^{free}_{IR}.}
The extra $U(1)_X$ accidental symmetry of the IR free confined scenario leads to a reduction of
$a$; this is because $R^{(0)}(X)=12/5$, which exceeds our conjectured upper bound $R^{(0)}(X)\leq 5/3$ for fields that can become free in the IR.   
So the conjectured diagnostic of this paper favors the interacting conformal phase scenario.

\centerline{\bf Acknowledgments}

I would like to thank Nati Seiberg and Steve Shenker for a discussion, last year,
which stimulated the conjecture explored here.    I'd also like to thank Nati for a helpful comment on the draft. This work was supported in part by UCSD grant DOE-FG03-97ER40546.  I would like to thank the IHES in Bures-sur-Yvette, where most of this work was
done, for  their hospitality and support. The final stage
of this paper was completed at the IAS, and I would like to thank the IAS, and the Einstein Fund, for their hospitality and support.

\listrefs\end